# Experimental demonstration and modeling of near-infrared nonlinear third-order triple-photon generation stimulated over one mode

**J. BERTRAND,[1] V. BOUTOU,[1] C. FELIX,[1] D. JEGOUSO,[1] AND B. BOULANGER[1,*]**

[1]*Université Grenoble Alpes, CNRS, Grenoble INP, Institut Néel, 38000 Grenoble, France*
*\*benoit.boulanger@neel.cnrs.fr*

**Abstract:**
Triple Photon Generation (TPG) is a third-order nonlinear optical interaction in which a photon of energy ℏω$_p$ splits into three photons at ℏω$_1$, ℏω$_2$ and ℏω$_3$, with ℏω$_p$ = ℏω$_1$ + ℏω$_2$ + ℏω$_3$. The triplets possess different quantum signatures from those of photon pairs, with a strong interest in quantum information. In the present study, we performed the first experimental demonstration of TPG stimulated over one mode of the triplet, at ℏω$_1$, the previous work on TPG concerning stimulation over two modes, at ℏω$_2$ and ℏω$_3$. The nonlinear medium is a KTiOPO$_4$ crystal pumped in the picosecond regime (15 *ps*, 10 *Hz*) at λ$_p$ = 532 nm. The stimulation beam is emitted by a tunable optical parametric generator: the phase-matching was found at a stimulation wavelength λ$_1$ = 1491 nm, the other two modes of the triplet being at λ$_2$ = λ$_3$ = 1654 nm in orthogonal polarizations. Using superconducting nanowires single photon detectors, the measurement of the polarizations and wavelength signatures of the two generated modes are in full agreement with calculations. It has been possible to generate a total number of photons *per* pulse on modes 2 and 3 up to $2 \times 10^4$, which corresponds to the generation of $10^4$ triplets *per* pulse, or $10^5$ triplets *per* second since the repetition rate is equal to 10 Hz. We interpreted these results in the framework of a model we developed on the basis of the nonlinear momentum operator in the Heisenberg representation under the undepleted pump and stimulation approximation.

## I. INTRODUCTION

Third-order parametric down-conversion, also called triple photon generation (TPG) is a nonlinear interaction consisting in the scission of a high energy pump photon $\hbar\omega_p$ into three lower energy photons ($\hbar\omega_1, \hbar\omega_2, \hbar\omega_3$). These three down-converted photons provide a new exotic quantum state of light that exhibits statistics going beyond the usual Gaussian statistics associated with coherent sources and optical parametric twin-photon generators. Actually, the direct simultaneous birth of three photons from a single one is indeed at the origin of intrinsic three-body quantum properties such as three-particle Greenberger-Horne-Zeilinger (GHZ) quantum entanglement [1] and Wigner functions presenting quantum interferences and negativities [2], [3] [4], [5], [6]. It may also open new horizons in quantum information, as for example the generation of heralded two photons states, which can be used in a qubit amplifier or in a device independent quantum key distribution protocol [7]. The first experimental demonstration of a TPG was done in the bi-stimulated regime [8], where additional coherent and intense stimulation beams at $\hbar\omega_2$ and $\hbar\omega_3$ were injected in a phase-matched bulk KTiOPO$_4$ (KTP) crystal while being pumped at $\hbar\omega_p$. This bi-stimulated configuration strongly increases the conversion efficiency of the down-conversion process and exhibits non intuitive continuous variable entanglement properties: actually, the tripartite entanglement is predicted to increase with seeding field for the two and three modes seeding. [9]. Nonetheless, the quantum properties of such a "mixed" state are difficult to implement in realistic quantum information protocols. The spontaneous TPG would offer the most interesting quantum state, but the conversion efficiency of such processes is very low, as shown by the quantum modeling in the weak interaction approximation in the Heisenberg representation [5]. However, an experimental demonstration of such a process was done in the GHz range [10], but the optical range, in particular in the Telecom range around 1500 nm, is still a challenge. With regard to this last objective, several unsuccessful attempts have been based on strategies using optical fibers [11], [12] or crystal waveguides [6] in order to boost the generation efficiency. An important step in this quest is the mono-stimulated TPG, in which a single stimulation beam, *i.e.* over a single of the triplet, is injected in the nonlinear medium in addition to the pump beam. It is the framework of the present paper where we report the first experimental demonstration of a mono-stimulated TPG in the picosecond regime at telecom wavelengths using a phase-matched KTP crystal. Using superconducting nanowires single photon detectors (SNSPD), the energy of the non-seeded modes ℏω$_2$ and ℏω$_3$ are measured and compared to a quantum model we developed on the basis of the nonlinear momentum operator in the Heisenberg representation.

## II. TPG CONFIGURATION

The mono-stimulated TPG is performed in the same nonlinear medium than that previously used for the bi-stimulated TPG [8]: it is a 1cm-long KTP crystal cut along the x-axis of the dielectric frame (x, y, z) and allowing birefringence phase-matching with the configuration of polarization described in Figure 1.

Then the energy conservation and momentum conservation write:

$$\Delta E = \hbar\omega_p - \hbar\omega_1 - \hbar\omega_2 - \hbar\omega_3 = 0 \quad (1)$$

$$\overrightarrow{\Delta k}(\omega_p, \omega_1, \omega_2, \omega_3) = \quad (2)$$
$$\frac{\omega_p}{c}n^-(\omega_p, \overrightarrow{u_p})\overrightarrow{u_p} - \frac{\omega_1}{c}n^+(\omega_1, \overrightarrow{u_1})\overrightarrow{u_1} -$$
$$\frac{\omega_2}{c}n^-(\omega_2, \overrightarrow{u_2})\overrightarrow{u_2} - \frac{\omega_3}{c}n^+(\omega_3, \overrightarrow{u_3})\overrightarrow{u_3} = \vec{0}$$

$n^-(\omega_m, \overrightarrow{u_m})$ and $n^+(\omega_m, \overrightarrow{u_m})$ are the eigen refractive indices at the circular frequency $\omega_m$ for a propagation in the direction $\overrightarrow{u_m}$ of the crystal, with m ≡ (p,1,2,3) [13]. Note that Eqs. (1)-(2) are valid for both the bi-stimulated and mono-stimulated TPG as well as for the spontaneous TPG. The full collinearity, *i.e.* $\overrightarrow{u_p} = \overrightarrow{u_1} = \overrightarrow{u_2} = \overrightarrow{u_3}$ is possible for both configurations of TPG, but the generation also occur under a given level of non-collinearity in the form of a cone in the mono-stimulated case, i.e. $\vec{u_p} = \vec{u_1} \neq \vec{u_2} \neq \vec{u_3}$.

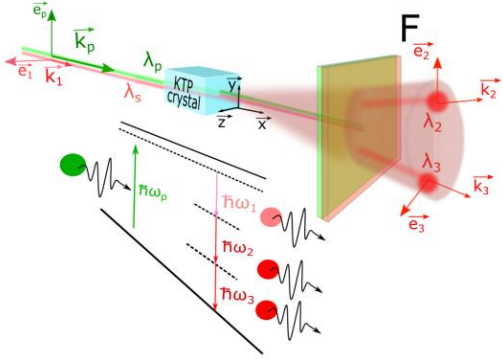

Figure 1: Schematic view of a TPG stimulated over one triplet mode, at $\lambda_1$, the two other triplet modes being at $\lambda_2 = \lambda_3$. The pump beam is at $\lambda_p$. The nonlinear medium is a phase-matched x-cut KTP crystal where (x,y,z) is the dielectric frame. $\vec{e_p}, \vec{e_1}, \vec{e_2}, \vec{e_3}$ are the unit electric field vectors of the interacting photons. $\vec{k_p}, \vec{k_1}, \vec{k_2}, \vec{k_3}$ are the corresponding wave vectors. $\chi^{(3)}_{yzzy}(\omega_p = \omega_1 + \omega_2 + \omega_3)$ is the relevant coefficient of the third-order electric susceptibility tensor of KTP. The pump and stimulation beams are filtered by a set of filters (F).

### III. EXPERIMENTAL SETUP

The experimental setup for mono-stimulated TPG is described in Figure 2. A 1064-nm laser beam with a FWHM pulse duration of $15\ ps$ at the repetition rate of $10\ Hz$ is splitted in two parallel beamlines by use of a beam-splitter (BS). The first beamline produces the pump photons at $\lambda_p = 532\ nm$ thanks to a second-harmonic generator based on a KTP crystal (SHG-KTP) cut at $\phi = 23°$ in the xy-plane. The second beamline is devoted to the pumping of a tunable-wavelength optical parametric generator (OPG - TOPAS Light Conversion) used for the stimulation of the TPG at $\lambda_1 = 1491\ nm$.

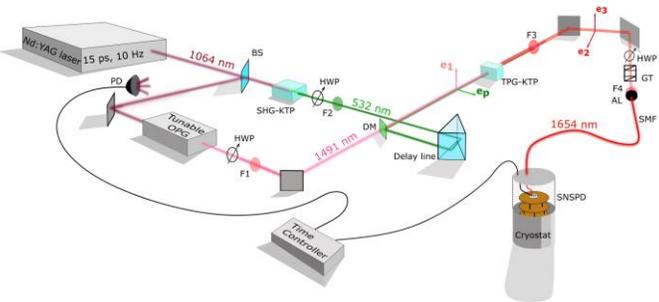

Figure 2: Experimental scheme for mono-stimulated TPG in a bulk KTP crystal using different optical components: a beam splitter (BS), an Optical parametric Generator (OPG), filters (F1, F2, F3, F4), a Second-Harmonic Generator (SHG-KTP), Half-Wave Plates (HWP), a dichroic mirror (DM), a Glan-Taylor prism (GT), a single-mode fiber (SMF), an aspheric lens (AL), a photodiode (PD) connected to a time controller, and a superconducting nanowire single photon detector (SNSPD) at 0.8 K in a cryostat.

We consider a semi-degeneracy for modes 2 and 3, i.e. $\lambda_2 = \lambda_3$, as justified in section IV. In order to avoid parasitic photons, the incoming beams are spectrally cleaned by two set of filters: (F1) for the stimulation and (F2) for the pump. The space and time overlap between the pump and stimulation beams is achieved using a delay line and a dichroic mirror highly reflective at 532 nm (DM). A telescope on the OPG output line is dimensioned so that the size of the stimulation beam waist *radius* in the crystal is $w_0^s = 150\ \mu m$, which is bigger than that of the pump beam measured at $w_0^p = 82\ \mu m$. The TPG is performed in a 1-cm-long KTP crystal (TPG-KTP). Another set of filters (F3), including notch and long-pass filters, is used to discriminate the modes 2 and 3 from the other sources of light. A bandpass filter FBH1650-12 (F4) at $1650 \pm 6\ nm\ (OD > 5)$ has a broad blocking region (200-1800 nm) and is directly posted close to the injection aspheric lens to filter any residuals parasitic photons from the crystal as well as pulsed ambient light. The detection is done on modes 2 and 3 using MoSi SNSPD from IDQuantique cool down to 0.8K in a liquid Helium cryostat. Because of the weakness of the signal, it is not possible to directly optimize its injection into the single mode fiber (SMF). It is why another parametric signal arising from the difference frequency generation between the pump and the stimulation beams $(1/532 - 1/1491 \rightarrow 1/827\ nm)$ is used prior to insert (F4) and (F3). A polarization selector composed of a Glan-Taylor prism (GT) and a Half-Wave-Plate (HWP) allow modes 2 or 3 of the triplet to be selected. The SNSPD is connected to a time-controller that is triggered by the signal from a silicon photodiode Thorlabs DET 110 (PD) installed on a pump laser loss. The triggering creates a temporal filtering that removes the continuous ambient light contribution. The quantitative measurement of the triple photon flux having to be performed at a weak level, typically $n_{triplets} \approx 0.01$ *photons per pulse*, calibrated neutral densities are used. This guarantees the statistical impossibility that photons coming from two or more triplets be present in the same detected pulse. This is mandatory to avoid any problems related to the very small time-width of the optical pulse (15 ps) compared to the recovery time of the detector (30 ns). The detection transfer function after the KTP crystal is estimated at $4.10^{-5}$. This extremely low value is mainly due to losses in the signal injection into the single mode fiber (SMF). The low quality of the beam profile does not fit well the SMF guided mode and reduces significantly the injection efficiency. The beam spatial jitter also reduces the average injection coefficient and increases its standard deviation.

### IV. MEASUREMENTS

#### A. Wavelengths, polarizations and divergence

We have chosen to perform the mono-stimulated TPG for which the two non-seeded modes are at the same wavelength, i.e. $\lambda_2 = \lambda_3$: the goal is to be in the same configuration than for the bi-stimulated TPG studied previously and in the same direction of propagation, i.e. along the x-axis [8]. Actually, the interest of such a phase-matching direction is that *(i)* the triplet wavelengths lye in the Telecom range, and *(ii)* the walk-off angle is null. Thus, a careful phase-matching investigation has been carried out in the bi-stimulated case, which also allowed us to select the relevant dispersion equations of the refractive indices required for the quantum model developed in section 0. For that purpose, the OPG has been tuned from $\lambda_2 = \lambda_3 = 1600\ nm$ to $\lambda_2 = \lambda_3 = 2000\ nm$ and the signal at $\lambda_1$ has been collected on a spectrometer. The measurement was investigated for different angles of propagation $\theta$ contained in the xz-plane. The result is shown in Figure 3.

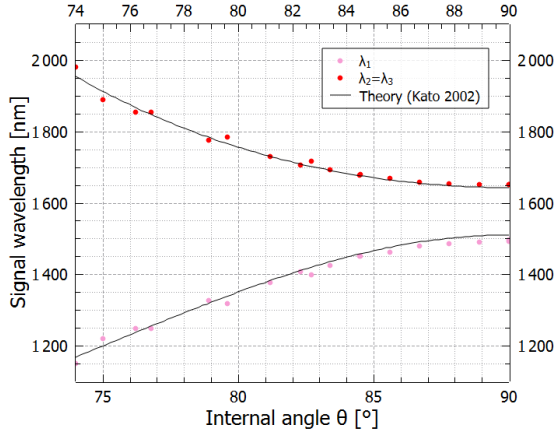

Figure 3: Measured and calculated phase-matching wavelengths of bi-stimulated TPG in the xz-plane of KTP pumped at $\lambda_p = 532\ nm$ as a function of the angle of spherical coordinate $\theta$ of the TPG phase-matching direction. $\theta = 0°$ corresponds to the z-axis and $\theta = 90°$ to the x-axis.

The experimental phase-matching wavelengths are in good agreement with the calculation using Eq. (2) and the Sellmeier equations of reference [14]. For the following, the mono-stimulated TPG will be performed along the x-axis ($\theta = 90°$) with the following set of wavelengths and polarizations : $(\lambda_p = 532\ nm,\ y - polarization)$ ; $(\lambda_1 = 1491\ nm, z - polarization)$ ; $(\lambda_2 = \lambda_3 = 1654\ nm,\ y - and\ z - polarizations)$. Then the corresponding effective coefficient is $\chi_{yzzy}^{(3)}(1654\ nm) = 7,8.10^{-22}\ m^2 V^{-2}$ deduced by using Miller's rule from $\chi_{yzzy}^{(3)}(539\ nm) = 14,6.10^{-22}\ m^2 V^{-2}$ measured in a previous work [15].

A clear signature of phase-matching for the mono-stimulated TPG is brought by Figure 5 where the number of generated photons in modes 2 and 3 are plotted as a function of the detuning of the polarization angle of the pump as well as of the signal. Figure 4 well shows that the signal at $\lambda_2 = \lambda_3$ drops to 0 when the phase-matching polarization conditions are switched off, i.e. $\lambda_p$ polarized along the z − axis or $\lambda_1$ along the y-axis, while it reaches a maximum when the phase matching conditions are switched on, i.e. $\lambda_p$ polarized along the y − axis and $\lambda_1$ along the y-axis.

By using several band-pass filters we determined that the peak wavelength of modes 2 and 3, which corresponds to the colinear contribution, i.e. $\overrightarrow{u_p} = \overrightarrow{u_1} = \overrightarrow{u_2} = \overrightarrow{u_3}$ in Eq. (2), are $\lambda_2 = \lambda_3 =$1654 nm as expected.

We also measured the divergency of the pump, stimulation and generated beam from the knife method, and we found a clear evidence of the generation over a cone. Actually, the divergence of modes 2 and 3 is 12.0 mrad, which is one order of magnitude higher than for the pump (2.1 mrad) and the stimulation (4.7 mrad).

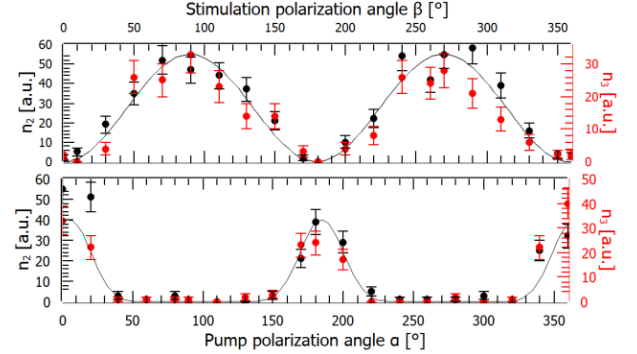

Figure 4: Number of photons $n_2$ and $n_3$ in modes 2 and 3 as a function of the polarizations of the pump and stimulation angles, ($\alpha$) and ($\beta$) respectively. 0° (resp. 90°) corresponds to the y-axis (resp. z-axis) of the dielectric frame of the KTP crystal. The pump and stimulation energies are $E_p = 26\ \mu J$ and $E_s = 21\ \mu J$, respectively.

## B. Numbers of photons *per* pulse

Figure 5 gives the number of photons *per* pulse on modes 2 and 3 as a function of the stimulation energy ranging from $62\ nJ$ to $21\ \mu J$ and for a fixed pump energy $E_p = 26\ \mu J$. The energies are controlled by calibrated neutral densities. Figure 5 shows a linear behaviour which will be interpreted in section 0. It has been possible to generate a total number of photons *per* pulse on modes 2 and 3, i.e. $n_2 + n_3$, up to $2 \times 10^4$. That corresponds to the generation of $n_{triplets} = 10^4$ *per* pulse, which gives $10^5$ triplets *per* second since the repetition rate is equal to 10 Hz. The corresponding quantum efficiencies are $\eta = \frac{n_{triplets}}{n_p} = 0.8 \times 10^{-11}$ and $\eta/n_1 = \frac{n_{triplets}}{n_p n_1} = 3.1 \times 10^{-24}\ Hz^{-1}$, where $n_p$ and $n_1$ are the numbers of pump and stimulation photons *per* pulse.

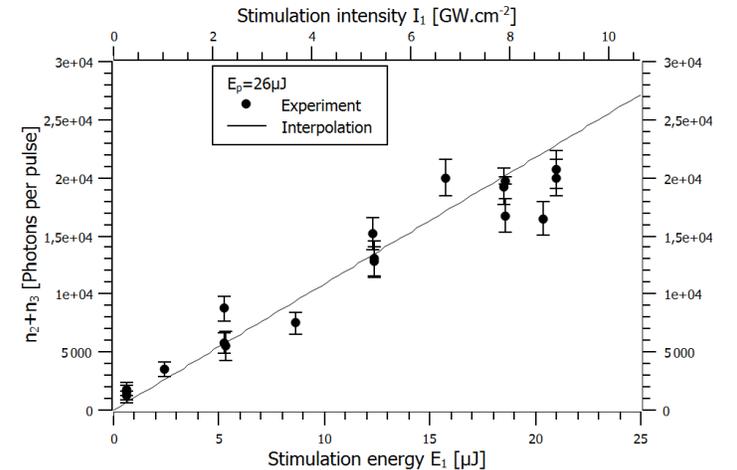

Figure 5: Measurements of the number of photons on modes 2 and 3 generated by mono-stimulated TPG in a 1-cm-long KTP crystal as a function of the stimulation Energy/Intensity at a given pump energy of $E_p = 26\ \mu J$.

## V. MODELING

We used a model in the Heisenberg representation based on the nonlinear momentum operator [5], [16], [17]. The propagation of the electromagnetic field in a nonlinear medium is then described by the following equation:

$$\frac{\partial a_j(\omega,Z)}{\partial Z} = -\frac{i}{\hbar}\left[a_j(\omega,Z), G_{nl}^{(3)}\right] \quad (3)$$

$G_{nl}^{(3)}$ is the third-order nonlinear momentum operator and $Z$ the space coordinate along the direction of propagation. The quantum fields are described in term of space dependant spectral mode operators $a_j(\omega,Z)$, with $j = (p,1,2,3)$, [16], [18], and the nonlinear process is considered as collinear, which means that Eq. (2) is reduced to a scalar equation, i.e.:

$$\Delta k(\omega_p, \omega_1, \omega_2, \omega_3) = \frac{\omega_p}{c}n_y(\omega_p, \vec{u}) - \frac{\omega_1}{c}n_z(\omega_1, \vec{u}) - \frac{\omega_2}{c}n_y(\omega_2, \vec{u}) - \frac{\omega_3}{c}n_z(\omega_3, \vec{u}) = 0 \quad (4)$$

where $\vec{u}$ is collinear to the x-axis of KTP in the present study, so that $n^- = n_y$ and $n^+ = n_z$.

A single collinearly propagating spatial mode of the electromagnetic field is then considered, keeping the same logic as in [16], which gives:

$$E(t,Z) = i\int_0^{+\infty} d\omega \sqrt{\frac{\hbar\omega}{4\pi c\epsilon_0 S}} a(\omega,Z) e^{-i\omega(t-\frac{Z}{c})} + H.C. \quad (5)$$

This description of a 1D electromagnetic field propagating along a single spatial direction is obtained by introducing the S area parameter. This area is defined by the cavity size in the two others spatial directions [18]. The nonlinear momentum model was developed from this 1D electromagnetic field formalism for both second-order and third-order spontaneous parametric down-conversion [5], [16], [17]. Mono-stimulated TPG was partially described [5], the pump field being assumed to be intense and non-depleted, so that the associated annihilation operator $a_p(\omega,Z)$ is well modelled by a classical spectral amplitude $A_p(\omega,Z=0)$ such that $|A_p(\omega,Z=0)|^2 = \langle n_p(\omega,Z=0) \rangle$. The number of pump photon is then given by $\int_\omega |A_p(\omega,Z=0)|^2 d\omega = n_p(0)$ where $n_p(0)$ is expressed in photon flux unit. Moreover, regarding the weakness of the magnitude of $\chi^{(3)}$, an approximation of weak coupling was made, expressed as $|\Gamma A_p| \ll 1$ where $\Gamma$ is proportional to $\chi^{(3)}$ [5]. The expressions of the field operators $a_1(\omega,Z)$, $a_2(\omega,Z)$ and $a_3(\omega,Z)$ are calculated through a solving of the space propagation equation (3).

Here we derived the third-order nonlinear momentum operator in the case of the mono-stimulated TPG in the general case, i.e. without the weak-coupling approximation. The stimulated mode, i.e. mode 1 in our case without loss of generality, is assumed to be intense and non-depleted so that $a_1(\omega,Z) = a_1(\omega,Z=0) = A_1(\omega,Z=0)$, where $A_1(\omega,Z=0)$ is the spectral amplitude of mode 1 at the entrance of the crystal. In these conditions, the third-order nonlinear momentum operator writes:

$$G_{nl}^{(3)}(Z) \quad (6)$$
$$\approx \int_0^{+\infty} d\omega_p \int_0^{+\infty} d\omega_1 \int_0^{+\infty} d\omega_2 \ \hbar\Gamma(\omega_p,\omega_1,\omega_2)\left(A_s(\omega_1,Z=0)a_2(\omega_2,Z)a_3(\omega_p - \omega_1 - \omega_2, Z) A_p^\dagger(\omega_p,Z=0)e^{-i\Delta k(\omega_p,\omega_1,\omega_2)Z} + H.C.\right)$$

According to Eqs. (3)-(5), it comes for the photon-flux spectral densities, i.e. the numbers of photons *per* pulse *per* Hz, on modes 2 and 3:

$$n_2(\omega,Z) = n_3(\omega_p - \omega_1 - \omega, Z) \quad (7)$$
$$= \begin{cases} 2\pi \dfrac{I_1(0)I_p(0)f^{(3)}(\omega)(\chi^{(3)})^2}{|C^{(3)}(\omega)|}\sin^2\left(\sqrt{|C^{(3)}(\omega)|}Z\right) \\ \quad \text{if } C^{(3)}(\omega) < 0 \\[6pt] 2\pi \dfrac{I_1(0)I_p(0)f^{(3)}(\omega)(\chi^{(3)})^2}{C^{(3)}(\omega)}\sinh^2\left(\sqrt{C^{(3)}(\omega)}Z\right) \\ \quad \text{if } C^{(3)}(\omega) > 0 \end{cases}$$

$I_p(0)$ and $I_1(0)$ are the pump and stimulation intensities at the entrance of the crystal, and the quantities $C^{(3)}(\omega)$ and $f^{(3)}(\omega)$ are defined by:

$$C^{(3)}(\omega) = 4\pi^2 I_1(0)I_p(0)f^{(3)}(\omega)(\chi^{(3)})^2 - \frac{\Delta k(\omega)^2}{4} \quad (8)$$

$$f^{(3)}(\omega) \quad (9)$$
$$= \frac{1}{(8\pi c^2 \epsilon_0)^2} \frac{\omega(\omega_p - \omega_1 - \omega)}{n_z(\omega_1)n_y(\omega)n_z(\omega_p - \omega_1 - \omega)n_y(\omega_p)}$$

It was admitted for the calculations above that the spectral bandwidth of modes 2 and 3 are wider than both pump and stimulation spectral bandwidths so that Eq.(4) writes: $\Delta k(\omega_p, \omega_1, \omega_2) = \frac{\omega_p}{c}n_y(\omega_p) - \frac{\omega_s}{c}n_z(\omega_1) - \frac{\omega_2}{c}n_y(\omega_2) - \frac{\omega_p-\omega_1-\omega_2}{c}n_z(\omega_p - \omega_1 - \omega_2) \approx \Delta k(\omega_2) \equiv \Delta k(\omega)$ since $\omega_2$ is taken equal to $\omega$. Furthermore, the third-order electric susceptibility $\chi^{(3)}$ has to be taken equal to $\chi_{yzzy}^{(3)}$, as defined in section IV.

Equations (7) show that the mono-stimulated TPG under the undepleted pump and stimulation exhibits two regimes: the weak coupling when $C^{(3)}(\omega) < 0$, i.e. for $4\pi^2 I_1(0)I_p(0)f^{(3)}(\omega)(\chi^{(3)})^2 \ll \frac{\Delta k(\omega)^2}{4}$, and the strong coupling when $C^{(3)}(\omega) > 0$, i.e. for $4\pi^2 I_1(0)I_p(0)f^{(3)}(\omega)(\chi^{(3)})^2 \gg \frac{\Delta k(\omega)^2}{4}$.

The integration of Eqs. (6) leads to the numbers of photons *per* pulse on modes 2 and 3, i.e.:

$$n_2(Z) = n_3(Z) = \int_\omega n_{2,3}(\omega,Z) d\omega \quad (10)$$

The numerical integration of Eq.(10) leads to a linear evolution of $n_{2,3}(Z)$ with the product $I_1(0)I_p(0)$ in the weak coupling regime, while it is an exponential behavior in the strong coupling regime. Looking Figure 5, it is obvious that the measured curve as a function of $I_1(0)$ for a fixed value of $I_p(0)$ is linear, which indicates that the experiments were performed in the weak coupling regime. We chose $\Delta k(\omega)$ as fitting parameter because the modelling was performed in the collinear approximation while the divergence relative to modes 2 and 3 is measured to be of about 4 times the divergence of the pump and stimulation beam, which may modify the "level" of momentum conservation. We then defined an effective phase-mismatch $\Delta k_{eff}(\omega) = \delta \Delta k(\omega)$ where $\delta$ is the fitting parameter. The interpolation is satisfying for $\delta = 2 \times 10^{-7}$ as shown in Fig. 5. It means that the effective mismatch is closer to 0 than the one given by the collinear model. An analytical expression of $n_{2,3}(Z)$ can be found thanks to the possible approximation of $\Delta k(\omega)$ as $\Delta k(\omega) \approx a + b\omega$, with $a = -3.3 \times 10^5 \ rad.m^{-1}$ and $b = 2.88 \times 10^{-10} \ m^{-1}.Hz^{-1}$, using the Sellmeier equations of reference [14]. Then it comes from Eqs. (7)-(10):

$$n_2(Z) = n_3(Z) \approx 4\pi^2 [f^{(3)}(\omega)]^2 \frac{I_p(0)I_1(0)\chi^{(3)^2}}{\delta|b|} Z \quad (11)$$

It is the first experimental validation of the third-order nonlinear momentum operator to the best of our knowledge.

## VI. CONCLUSION

In this paper, we report the first experimental demonstration of a mono-stimulated TPG in a 1-cm x-cut KTP crystal. The phase matching conditions were experimentally calibrated with a bi-stimulated TPG to perform a semi-degenerate mono-stimulated TPG. Both polarization and spectral signatures are in good agreement with momentum and energy conservations. It has been possible to generate up to $n_{triplets} = 10^5 \ s^{-1}$ in the Telecom range, the corresponding quantum efficiencies being $\eta = \frac{n_{triplets}}{n_p} = 0.8 \times 10^{-11}$ and $\eta/n_1 = \frac{n_{triplets}}{n_p n_1} = 3.1 \times 10^{-24} \ Hz^{-1}$, where $n_p$ and $n_1$ are the numbers of pump and stimulation photons. In order to interpret these results, we developed a quantum model on the basis of the nonlinear momentum operator in the Heisenberg representation under the undepleted pump and stimulation approximation. We identified two running regimes, the so-called weak- and strong-coupling, and showed that our experiments were performed under the weak-coupling regime, with a linear behavior of the generated triplets as function of the stimulation intensity. The agreement between experiment and theory is very good by introducing an effective phase-mismatch, which suggests to improve our model by taking into-account the non-collinearity of the mono-stimulated TPG. Another next step will be devoted to the quantum properties of the triplets generated by the mono-stimulated TPG. A first study could be the analysis of the three-photons coherence by observing the sum-frequency generation in a nonlinear crystal as previously done in the case of twin-photons generated by a second-order SPDC [19]. Previous calculations had shown that non-classical correlations can be put into light in this scheme. Actually, the summed-frequency fields show a fluorescence-like behavior in a very low-stimulation regime, but a classical-like behavior as soon as stimulation is high enough to be perceptible in the non-injected generated field spectrum [5].

## ACKNOWLEDGMENTS
We wish to acknowledge support from the Fédération QuantAlps for the PhD of Julien Bertrand.

## CONFLICT OF INTEREST
The authors have no conflicts to disclose.

## DATA AVAILABILITY STATEMENT
The data that support the findings of this study are available from the corresponding author upon reasonable request.